\documentclass{article}
\usepackage[T1]{fontenc} 
\usepackage[utf8]{inputenc} 
\usepackage{amsmath,cite,url}
\usepackage{graphicx}
\usepackage{color}
\usepackage{enumitem}

\usepackage{lineno}

\title{An analysis of Iranian Music Intervals based on Pitch Histogram}

\author
  {Sepideh Shafiei} 

\begin{document}

\maketitle
\begin{abstract}
\par Since the early twentieth century, intervals and tuning systems have been subjects of discussion among Iranian musicians and scholars. The process of Westernization and then a cultural back to roots movement are among the reasons that motivated debates about the appropriate tuning in this musical culture. In this paper, we first review the historical context of the intervals in Perso-Arabic musical culture since Fārābi in the tenth century. Then we focus on the audio histogram of the vocal performance of each piece in the repertoire (radif) of Karimi, one of the masters of the art, and use Dynamic Time Warping for alignment of pitch and MIDI notes. We collected the intervals used in the performance of each piece (gushe) in the repertoire and then analyzed the results. Unlike the traditional methods of measuring the frequency of each note played on the tār (an Iranian lute) practiced by contemporary music scholars, our computational method is independent of a given instrument and can be executed on any performance with minimum effort. 

\end{abstract}

\section{Introduction}

The intervals of Iranian music are not very fixed, and they are always a subject of discussion among musicians and Iranian musicologists. In this paper, we first review the historical background of the intervals in Persian classical music, then we use the histogram of audio to find the intervals in the vocal repertoire of  Karimi [1], one of the most prominent masters of the art in the twentieth century. In order to do so, we fitted the histogram mountain associated to each note with a Gaussian curve and found the peak. In some exceptional cases, we used degree-two polynomials to model the notes in audio. The study of intervals through analysis of the histogram of performance led us to categorize the types of histogram peaks. It also helped us observe which intervals are fixed and which intervals have more variations and speculate the reason behind these differences. Furthermore, we compare our results with the intervals measured by Talāi and Farhat and presented tables for comparison of contemporary Persian music intervals with those of the theories of Fārābi [2] and Marāghi [3]. 
\par The repertoire of Iranian classical music (\textit{radif}) consists of seven \textit{dastgāhs} and five \textit{āvāzes} (secondary dastgāhs). Each dastgāh consists of several pieces (\textit{gushes}). These gushes are in different maqāms  and they are related to each other through a special order, which provides a path for modulation from one maqām to another, inside a given dastgāh [4]. In many contemporary Persian musicological texts, the word “maqām” is used without a precise definition, and in some cases, the definition exists, but the usage of the word remains vague. Among the clear explanations and applications of this word, one can mention Alizādeh’s definition. In his view, maqām or mode is a scale in which certain tones have specific functions, and certain melody models might exist in pieces performed in that maqām. 

\section{Historical Overview}\label{sec:introduction}
\par Old Persian and Arabic music theory treatises, such as Fārābi’s Kitāb al-musiqi al-kabir in Arabic, Marāghi’s Jāme al-alhān in Persian, and Ormavi’s Kitāb al-advār in Arabic have received attention from Iranian ethnomusicologists and music scholars in the past decades. In this paper, we consider two of the most important historical sources for the theory of Perso-Arabic musical intervals during tenth to fourteenth century: Kitāb al-musiqi al-kabir by Fārābi (died 950), and Jāme al-alhān by Marāghi (died 1434). In the historical theories of music, they use the fractions of a whole length of a string to measure the intervals. For simplicity, I have converted the fractions to the cents system. The following formula gives the interval between a and b in cents.
\begin{equation}
\frac{a}{b}\rightarrow1200\times\log_2{\frac{a}{b}}
\end{equation}
For example, $256/243$ will be converted to $1200\times\log_2{\frac{256}{243}}\approx90$ cents. Fārābi introduces exactly 22 pitches in each octave [5]. Fārābi then explains the acceptable combinations of the pitches in an octave that can be used in a single melody. He explains that some of the pitches are compatible (\textit{mojānes}) with each other and some are not. Fārābi suggests that there are 7 pitches in each octave that form a pleasant scale. He mentions three scale types as follows 
\begin{enumerate}[label=\Roman*.]
\item 204, 408, 498, 702, 906, 996, 1200
\item 204, 355, 498, 702, 853, 996, 1200
\item 204, 303, 498, 702, 801, 996, 1200
\end{enumerate}
On each string, the second and fourth intervals are natural, 204, and 498 respectively, but the third interval can be 303 (\textit{vostā fars}), 355 (\textit{vostā zalzal}), or 408 cents.
The second treatise that we consider is Marāghi’s Jāme al alhān. There are three types of interval in Marāghi’s system: \textit{tanini} (=9/8~204 cents), \textit{baqieh} (=256/243~90 cents), and \textit{mojannab} which can be large \textit{mojannab}, 180 or 182 cents and small mojannab, 112 or 114 cents. Marāghi’s scale is similar to Pythagoras’s intervals, explained in Safi al-Din Ormavi’s treatise and can all be produced using the circle of fifths, with the fifth interval equal to 702 cents. Owen Wright links these numerical relationships to the Neo-Platonism of the period [6]. 
\par The musical intervals became a subject of study again in the twentieth century. Table \ref{tab} shows the comparison of Marāghi’s intervals to the current Persian scale, reported/measured by Farhat [7] and Talāi [8] respectively. Both Farhat and Talāi have measured the intervals between the frets of tār and setār. Farhat mentions that he has measured the intervals between the frets for two tārs and one setār using a stroboconn between 1959 and 1964. The perfect fourth and perfect fifth intervals in Farhat’s measurements seem to be taken from the equal temperament scale. As can be seen in Table \ref{tab}, all the natural notes, as well as some of the flat notes are approximately the same in both Farhat and Marāghi’s octave division. The differences happen mostly in neutral intervals and sometimes on the flat and sharp pitches. The letter “k” stands for koron, the half-flat sign suggested by Vaziri [9]. Farhat’s intervals have in general more similarities with Marāghi’s intervals. However, there is a big difference between Marāghi’s neutral intervals with both Farhat’s and Talāi’s measurements for the neutral intervals in contemporary music of Iran. Talāi’s intervals seem to have been approximated in a way that all the measurements are multiples of 10.

\begin{center}
\begin{table}
\centering
\footnotesize{
\begin{tabular}{ | c | c | c | c | }
\hline
Note Names &	Farhat &  Talāi & Marāghi \\
\hline
Sol &	0&	0 &	0\\
\hline
La b	&90	& - & 90\\
\hline
La k &135	&140	&180	\\
\hline
La& 205 & 200 &204	\\
\hline
Si b &295	&280	&294\\
\hline
Si k &340	&350	&384	\\
\hline
Si&410&380&408\\
\hline
Do&500&500&498\\
\hline
Re b	&565	&580	&588	\\
\hline
Re k &630	&640	&678	\\
\hline
Re &700 & 700&702	\\
\hline
Mi b & 790&- &	792 \\
\hline
Mi k	&835&840	&882	\\
\hline
Mi&905&900	&906	\\
\hline
Fa&995&980&996\\
\hline
-&1040&1050	&1086\\
\hline	
Fa\#  &1110&-&1176 \\
\hline
Sol&1200	&1200&1200\\
\hline
\end{tabular}
}
\caption{\label{tab}Intervals Comparison}

\end{table}
\end{center}

\section{Methodology}
\subsection{Pitch Recognition}
The first step of our computational process of finding intervals is pitch recognition. There are different algorithms for this purpose. All these algorithms use the fundamental frequency of the sound to quantify pitch. Fundamental frequency is the lowest frequency of the sound wave and corresponds to its most dominant perceived pitch. Monophonic voice pitch estimation algorithms have been evaluated and discussed extensively in MIR literature. Gomez et al. have evaluated both algorithms CREPE [10] and pYIN [11] as state-of-the-art pitch recognition methods for monophonic voice [12]. They have evaluated the results of both of these algorithms on iKala dataset and obtained almost similar results for both pYIN and CREPE: 91\% of Raw Pitch Accuracy for pYIN for monophonic voice and 90.5\% accuracy for CREPE. 
We compared the output of pYIN and CREPE for our data which is monophonic voice, by listening to the F0 traces of the fifteen gushes of shur from Karimi’s vocal radif and the ones from pYIN sounded more accurate.We used the pYIN as our pitch recognition algorithm with the following parameters: step size of 256, block size of 2048, low amplitude suppression of 0.1, onset sensitivity of 0.7, prune threshold of 0.1, and threshold distribution of 2. 

\subsection{Pitch Histogram and its Peaks}

\begin{figure}
 \centerline{\framebox{
 \includegraphics[width=0.9\columnwidth]{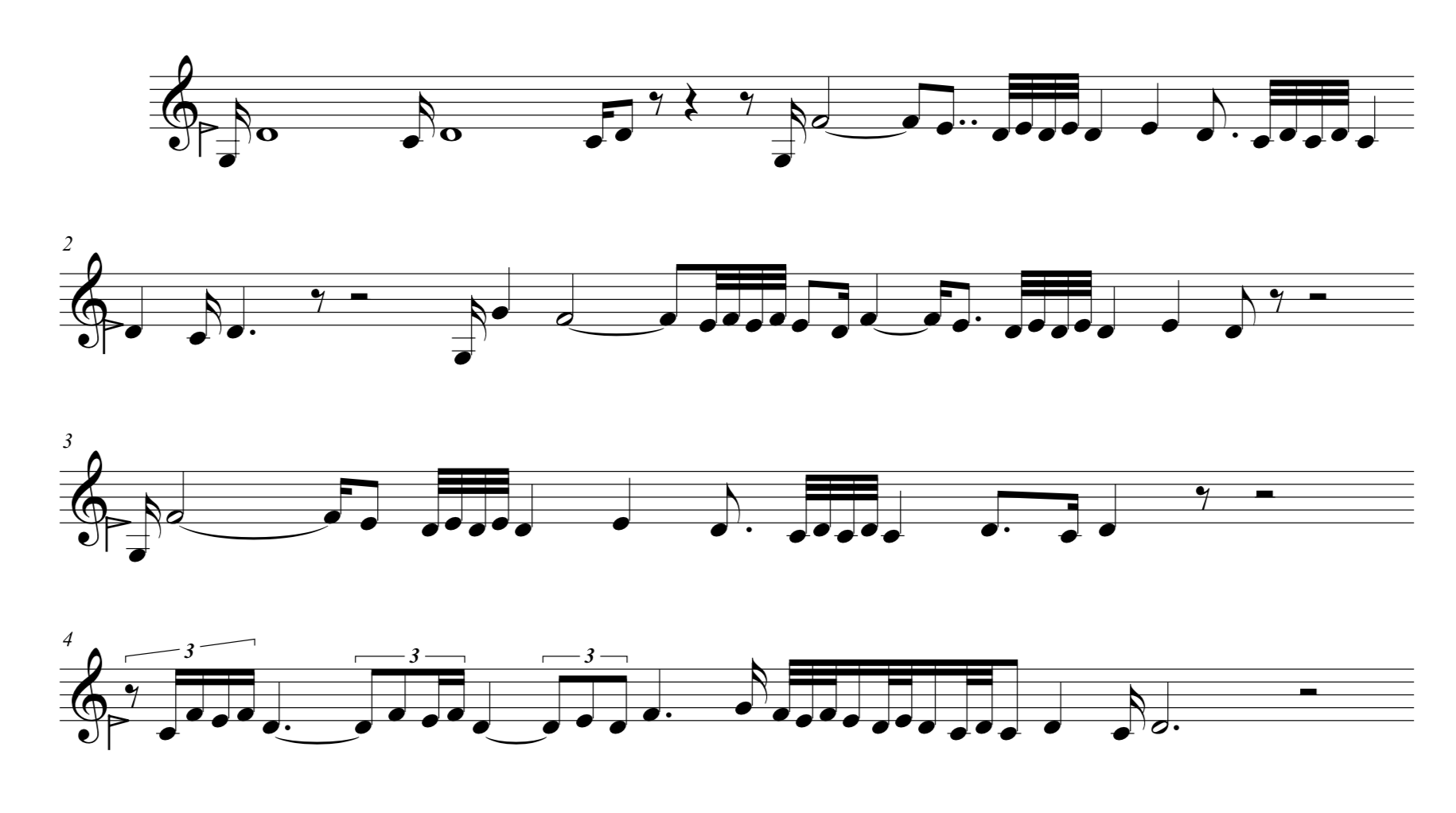}}}
 \caption{Transcription of the darāmad of shur}
 \label{fig:daramad}
\end{figure}

\begin{figure}
 \centerline{\framebox{
 \includegraphics[width=0.9\columnwidth]{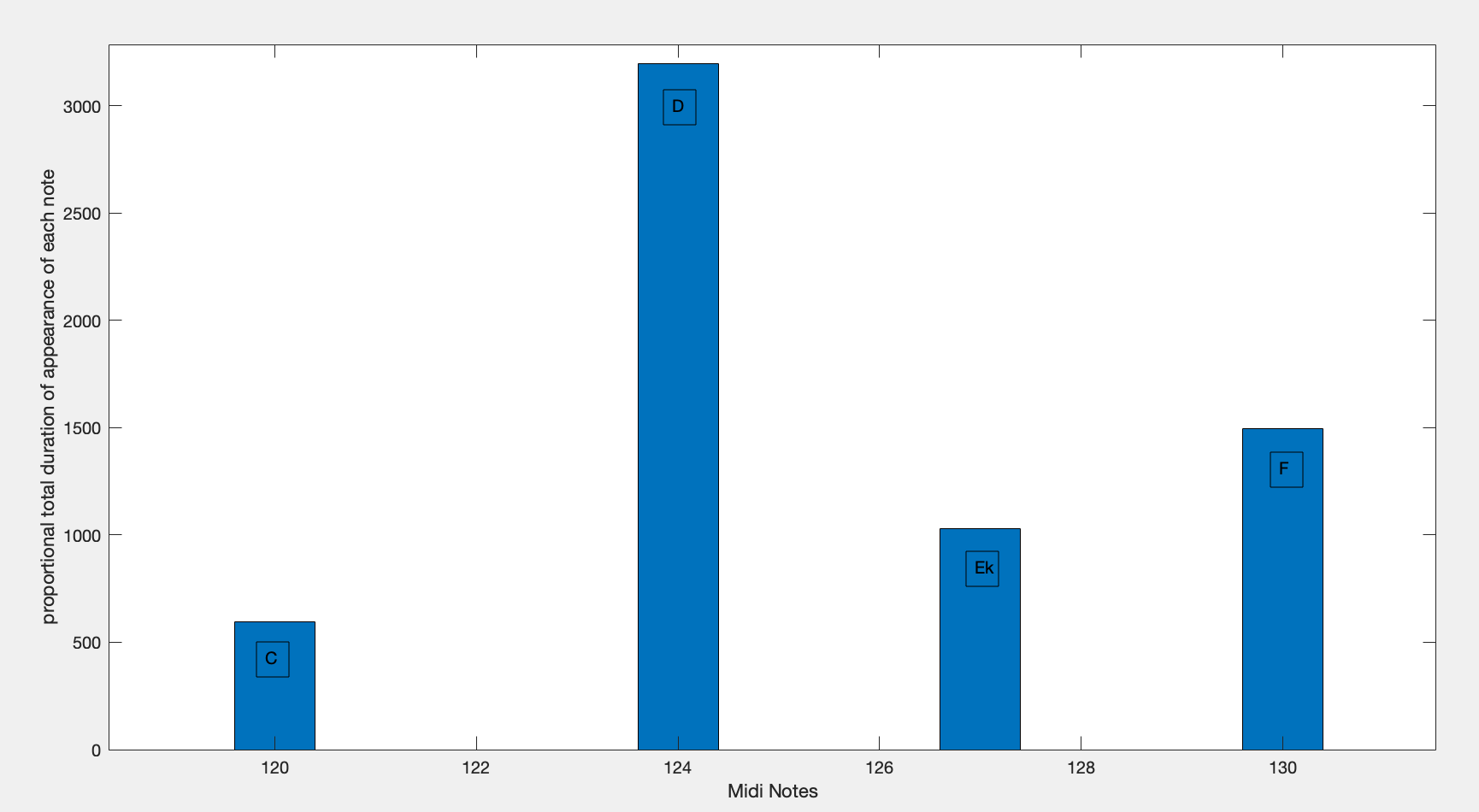}}}
 \caption{MIDI Histogram}
 \label{fig:discrete histo}
\end{figure}

Our computational analysis of the intervals is mainly based on histograms of the performed pitches. Pitch histogram has been used before by Bozkurt for analysis of Turkish Makam Music [13]. A discrete histogram is a bar chart in which the height of each bar represents the total duration of all occurrences of a note\footnote {In other words, a MIDI histogram can be viewed as the weighted frequency of occurrences of the notes, where the weight is the total duration of each note’s occurances.}. Figure  \ref{fig:daramad} shows the transcription of the first gushe of the vocal radif of Karimi, called darāmad of shur. In Figure \ref{fig:discrete histo} we see the MIDI histogram of the same piece. We have only included the notes which their total duration is more than 2 percent of the total duration of the whole piece. \footnote {For our purpose, we multiplied the MIDI numbers by two, so that we can represent neutral intervals as well.}

\begin{figure}
 \centerline{\framebox{
 \includegraphics[width=0.9\columnwidth]{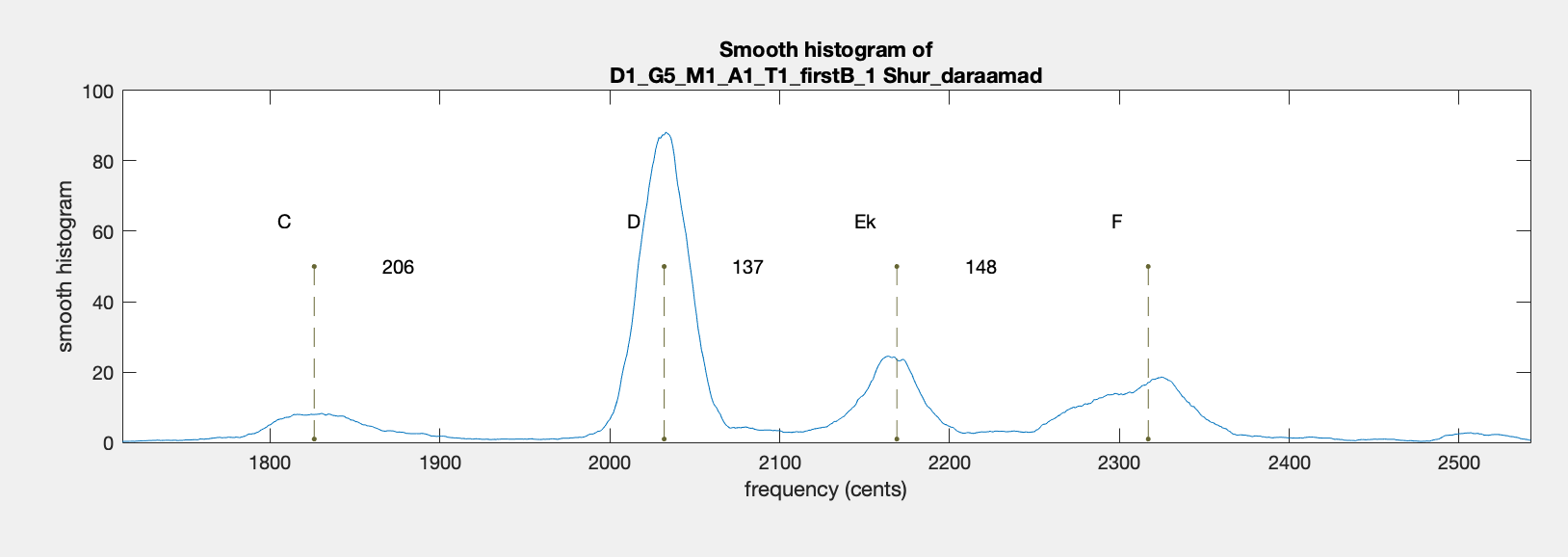}}}
 \caption{Audio Histogram}
 \label{fig:audio histo}
\end{figure}

Next we look at the same segment performed by Karimi, where each note has many adjacent frequencies as well. Hence the histogram of the audio recording of the same segment has a continuous mountain shape rather than a discrete bar chart. The performance does not need to match the notation exactly. Figures \ref{fig:audio histo} shows the output of our code for the audio recording of the darāmad of shur performed by Karimi. For finding the audio histogram, we find the total number of occurrences of each fundamental frequency. The reason that the mountain of Ek in the audio histogram is counterintuitively higher than F is that the range of the mountain of F in audio is wider than Ek. In audio, the area under each mountain is a better representative than the height of the mountain for finding the total duration of each note. We then smooth the histogram in order to find the exact peak of each mountain by finding the moving average to smooth out short-term fluctuations and highlight the general trend of the data and we get a semi-Gaussian  curve. The vertical dashed lines in Figure \ref{fig:audio histo} show the peak of each mountain. The first peak corresponds to C, the second peak to D, the third peak to E-koron and the last one to F. The numbers between the vertical lines show the interval between the notes in cents according to Karimi’s performance. The interval between C and D is 206 cents, the interval between D and E-koron is 137 cents and the interval between E-koron and F is 148 cents. The process of finding the peaks and intervals has been described below. In order to find a mapping from the audio to MIDI histogram, we use shāhed (literally witness)\footnote {According to Wright, shāhed is the most prominent pitch of the gushe, the salience of which is marked primarily by relative duration.} of each piece.  Intuitively, shāhed (witness note) of each gushe should be the highest peak of the histogram. This intuition works well in the case of the darāmad of shur. However, we also need to pay attention to the range of each peak, since in reality, what we need to find as shāhed is the note which has the highest area under its mountain range in the histogram, which mathematically is the definite integral of the histogram curve in the range of the mountain. Musically, this value represents the cumulative number of times that the sound frequencies in the range of each note are performed.
 \par
\begin{figure}
 \centerline{\framebox{
 \includegraphics[width=0.9\columnwidth]{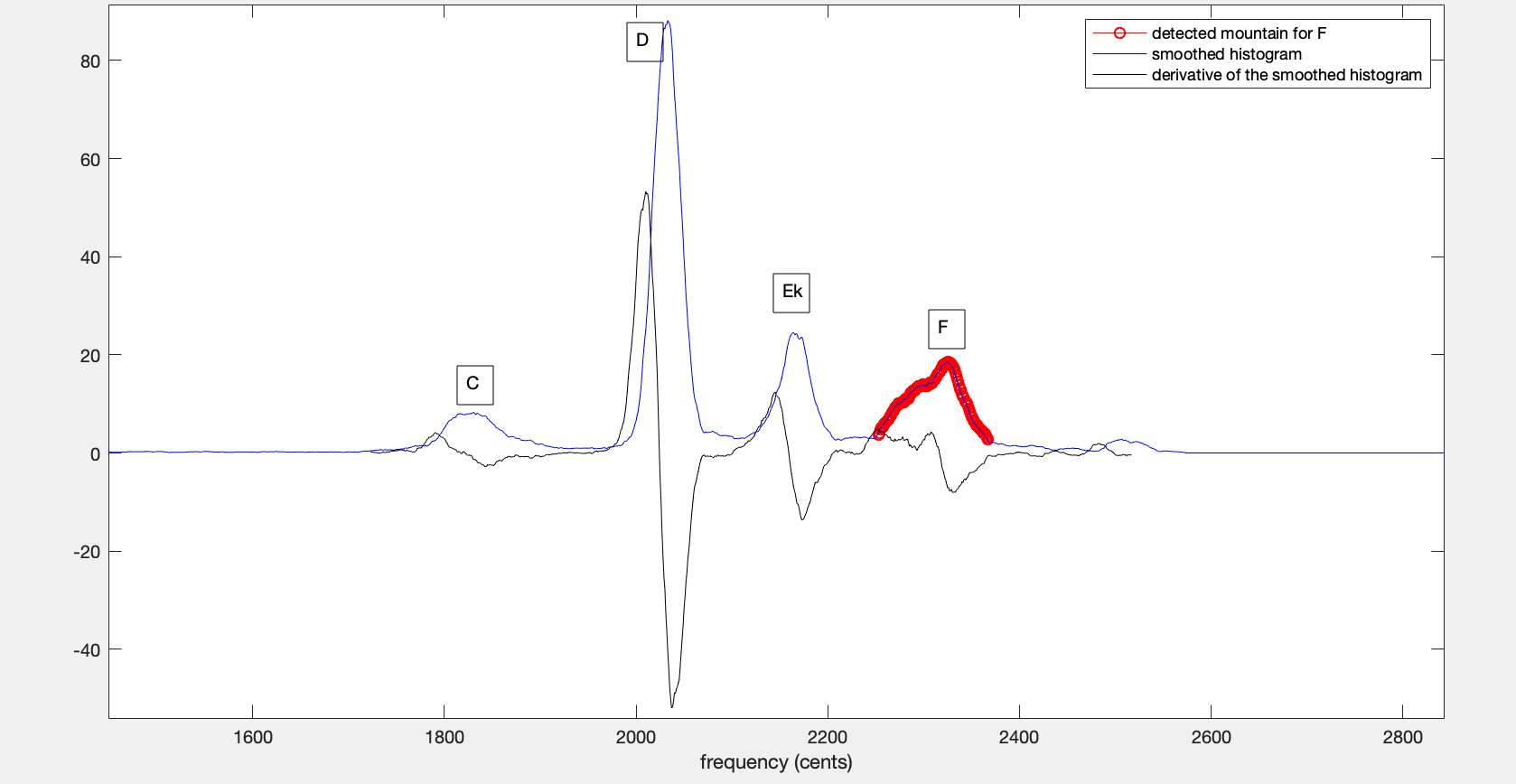}}}
 \caption{Range of F}
 \label{fig:F range}
\end{figure}

Finding the range of the mountain is itself a challenging task. We have used different methods based on the shape of the mountain and the shape of the derivative of the smoothed histogram. Following the sign changes in the derivative curve gives us an indication of the range of each peak. For example, Figure  \ref{fig:F range} shows the way we find the range of the mountain for the note F in darāmad of shur.

As can be observed from many audio histograms of different gushes in the repertoire, some notes are characteristically more flexible in terms of pitch. Finding the reason behind each relatively wide mountain requires a thorough study, which will be discussed in Section \ref{Types of Peaks}. After finding the range of each peak, we approximate the peak by a tilted Gaussian curve so that we can find a better peak. Choosing reasonable peaks is very crucial. These peaks are then used in finding the shāhed of each gushe in audio and then compare the shāhed of audio with the shāhed of MIDI, in order to find the scale automatically from audio. The amount of flexibility of each of these peaks are discussed later in Section 2.8. Here is the equation of the model function we use for a tilted Gaussian curve:
\begin{equation}
{y=c}_1+c_2x+c_3e^{-(x-c_4)^2/c_5}
\end{equation}
Then we use a non-linear curve fit function in MATLAB to find the parameters $c_1,\ldots,\ c_5$ so that the above equation fits our data.  After finding the peak for each mountain in the histogram of audio, we have the frequency of each note in the performance in cents, as a result, we also have the frequency of the shāhed for each gushe. In section \ref{Types of Peaks}, we will discuss different types of histogram peaks that we have observed in Karimi’s performance of radif.

\section{Typology of the Observed Peaks}
\label{Types of Peaks}

We use the peak of the audio histograms to find the performed musical intervals. In this process, it is crucial to consider the shapes of the mountains and their peaks for each note. These shapes are varied, and the difference between these shapes could give us some insight into the intervals and the notes performed in each piece in the repertoire. In our studies, we have noticed the following types of mountains in the histogram of voice for Karimi’s vocal radif.

\par
\begin{figure}
 \centerline{\framebox{
 \includegraphics[width=0.9\columnwidth]{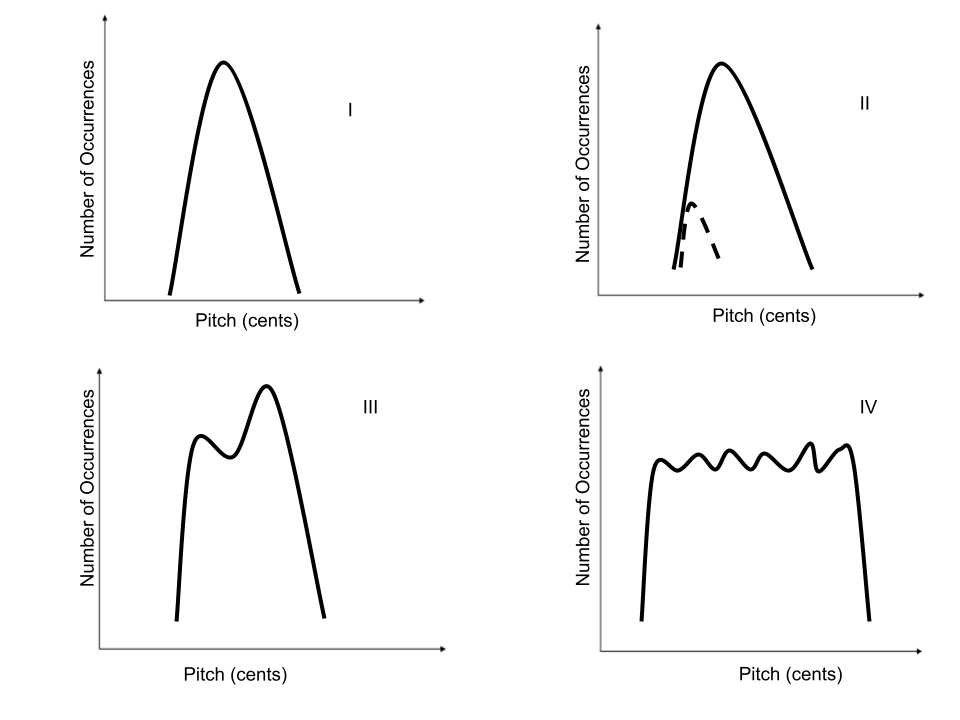}}}
 \caption{Types of Pitch Histogram Peaks}
 \label{fig:Peaks}
\end{figure}

 I. The ideal type of mountains can be fitted with a Gaussian function with a minimal error boundary. In this case, the peak is very clear and relatively easy to find, and in the comparison of MIDI and audio, the audio peak is clearly associated with one MIDI note. Figure  \ref{fig:Peaks}-I shows a general shape for this type of mountains.

\par II. The second type of mountain looks similar to the first type. The only difference is that when we compare the histograms of audio and MIDI, we notice two MIDI notes in the range of this mountain. The number of occurrences of one of the peaks is generally much lower than the prominent peak, their interval distance is about 50 cents or less, and the lower peak is hidden in the larger mountain. In this case, if we do not consider the transcription at all, we see a mountain that is very similar to the first type, but if we consider the transcription, after finding the histogram of each note separately, we find a small hidden mountain, inside the larger mountain. In some cases, this small mountain and peak causes a bump in the larger mountain and is more visible in the audio histogram. This type mostly happens when we have a variable note in the gushe. Figure  \ref{fig:Peaks}-II shows the shape of this type.

\par III. Sometimes a mountain has two distinct peaks. The height of these two peaks can be the same or different. In this case, we need to decide whether the whole mountain in audio histogram corresponds to one note, or two separate notes in the MIDI. If the two peaks correspond to two separate notes, then we need to find two different peaks. If the whole mountain corresponds to a single MIDI note, then we need to decide whether to take the higher peak or find a peak in the middle of the two using the Gaussian peak fit.

\par IV. The mountain shape looks flat and curvy on the top over a relatively large interval, sometimes about 30 cents or more. In this case, we need to find out if it corresponds to a single note or two separate notes and why the peak is so flat. Sometimes these kinds of peaks correspond to the notes, where the vocalists have performed vibrato on a note multiple times throughout the gushe. Figure  \ref{fig:Peaks}-IV shows the general shape of this type of mountains.

\par The other mountain shapes are a combination of these four types. In cases II, III, and IV, there is not a definite answer for the best peak. One can speculate case by case depending on the piece and the reasons that caused such mountain shapes. In Section \ref{noteHisto}, we propose a computational method for finding the peaks for each note, which relies on the alignment of audio and transcription. This new method will help us find the peaks of types II, III, and IV. With this information, we are ready to align the audio and MIDI in the next section. 

\section{Alignment of Pitch and MIDI}\label{alignment}

Parallel to the audio we have made a table corresponding to the MIDI file which contains the MIDI note and the duration of each note based on the transcription. The problem of MIDI to audio matching has a long history in the Music Information Retrieval community. There are different algorithms for comparing and matching MIDI and audio sequences. Ewert et al. mention three different approaches for the alignment problem: Dynamic Time Warping  (DTW), a recursive version of Smith-Waterman algorithm , and partial matching [14]. Based on our data, we decided to use the Dynamic Time Warping algorithm in MATLAB to compare MIDI and audio pitch curves. DTW is one of the most common algorithms that is used to compare two given time series in MIR and speech recognition [15].  
\par
\begin{figure}
 \centerline{\framebox{
 \includegraphics[width=0.9\columnwidth]{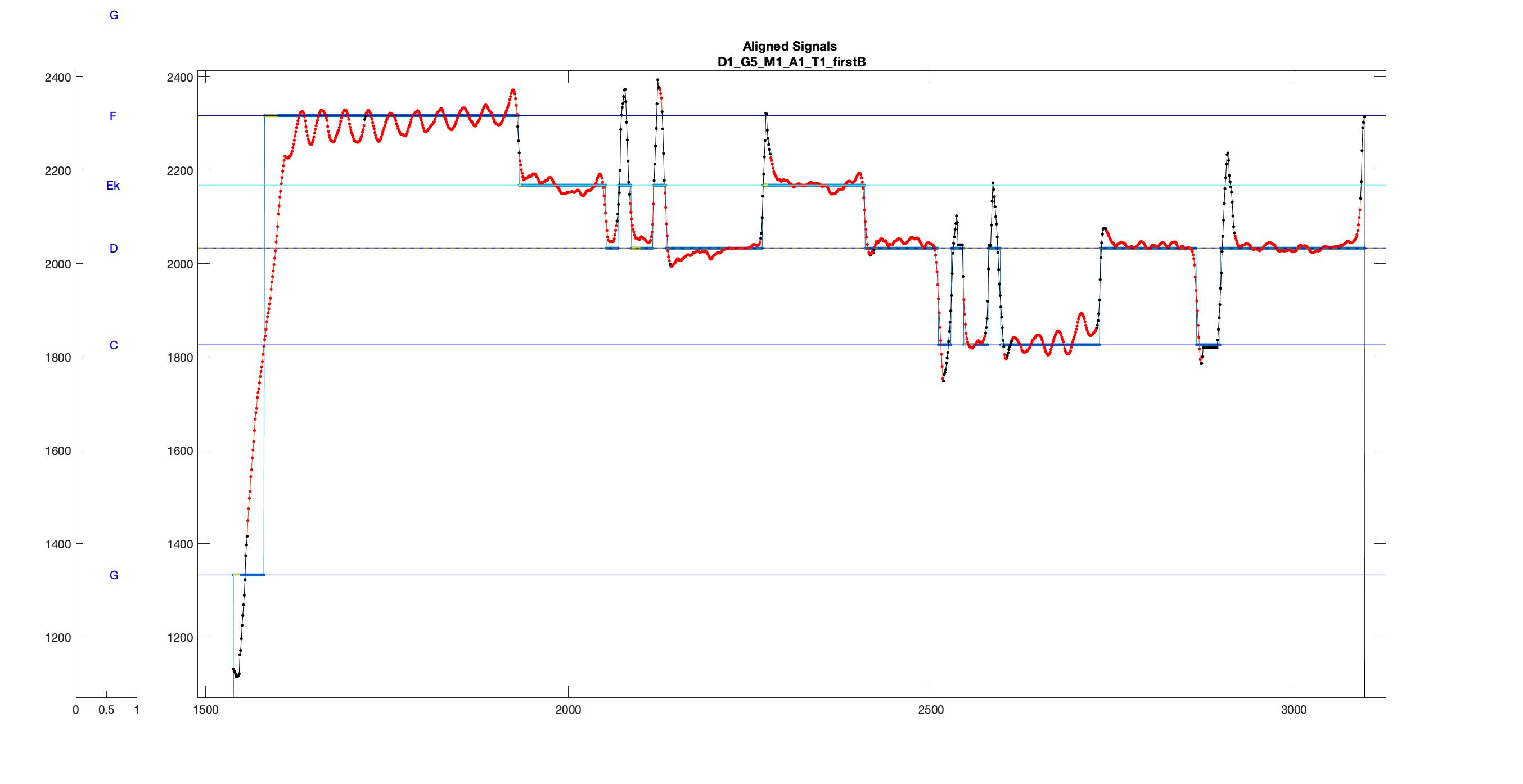}}}
 \caption{Aligned Pitch and MIDI}
 \label{fig:aligned}
\end{figure}

 For alignment of audio and MIDI, there are different approaches based on DTW. For our purpose, as in this paper we only deal with the monophonic voice, and the pYIN pitch recognition algorithm works very well and with minimal error on our data, we directly used DTW to align F0 with MIDI.  In order to evaluate the alignment process, we listened to the F0 playback and also used audacity to manually annotate the first three pieces of the repertoire. For the steady notes, the difference between the manually annotated ground truth and DTW alignments was at most 24 ms. For tekyes \footnote {tekye is a form of vocal ornamentation in Iranian vocal music which involves sudden jumps in frequency [16].}, the difference between the manual annotations and DTW markings was at most 54 ms. Plans for future work includes using the DTW-based alignment algorithm introduced in [12]. In order to compare the MIDI sequence of notes and the audio sequence of frequencies as two time series using the DTW algorithm, we need to prepare audio and MIDI as two comparable sequences. To do so, parallel to the audio sequence of pitches, we make a sequence for MIDI, in which we repeat the frequency associated with each MIDI note a number of times, so that the total length  of MIDI and audio sequences will be the same. DTW works very well for aligning Karimi’s vocal performance with Masudiye’s transcription. Figure  \ref{fig:aligned} shows the aligned MIDI and pitch diagrams for a sentence in the darāmd of shur.

\section{Histogram of Notes}\label{noteHisto}

\begin{figure}
 \centerline{\framebox{
 \includegraphics[width=0.9\columnwidth]{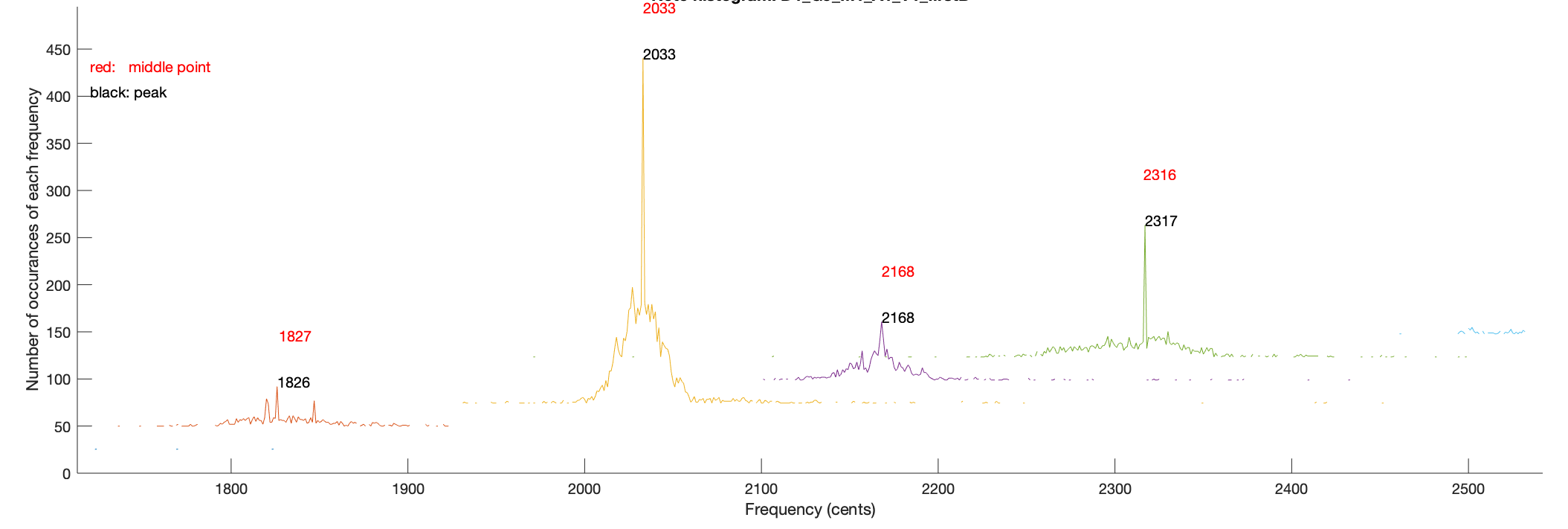}}}
 \caption{Note Histogram of the darāmad of shur}
 \label{fig:Notehisto}
\end{figure}

This method is developed as a way for verifying our results from finding the histogram peaks in the cases where we have unclear peaks in the audio histogram of a gushe. It can specifically help in the peaks of types II, III, and IV described in Section \ref{Types of Peaks}. This method relies on the alignment of the audio with the transcription. After matching MIDI and pitch in Section \ref{alignment}, we find the histogram of each note separately. In this process we have to find all the frequency points in the audio which are matched to a particular MIDI note, and then find the histogram of those points. Figure  \ref{fig:Notehisto} shows the histogram of each note for the darāmad of shur, performed by Karimi. The horizontal axis shows the frequency in cents, the vertical axis represents the number of occurrences of each fundamental frequency during the performance. Each of the mountains in this figure shows the frequencies that are matched to one MIDI note. 

\section{Results}\label{results}

\begin{figure}

 \centerline{\framebox{
 \includegraphics[width=0.39\columnwidth]{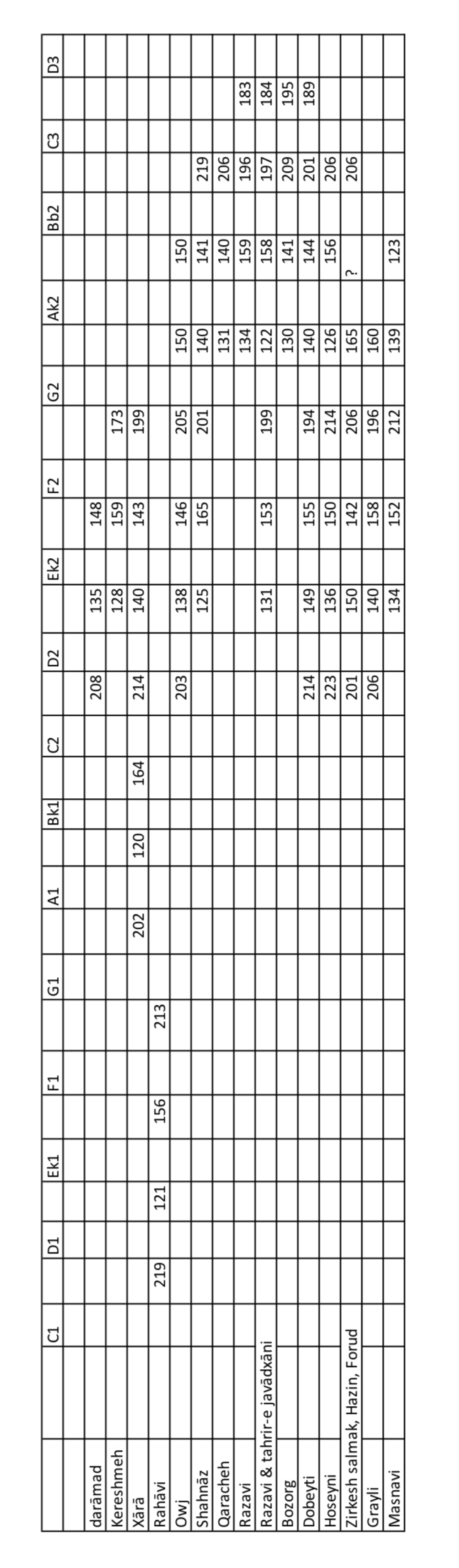}}}
 \caption{All the observed intervals in the dastgāh of shur}
 \label{fig:all intervals}
\end{figure}

 Figure \ref{fig:all intervals} shows all the intervals in the 15 gushes of shur. Table \ref{shur} represents the average of the detected intervals in Karimi's performance of the dastgāh of shur in comparison with Vaziri, Talai, and Farhat theories of the intervals in Persian classical music. We have eliminated the cases with very few samples in order to have a more reliable estimation.

\par To summarize, the analysis of the intervals of the fifteen gushes of shur indicates that the intervals performed in this dastgāh are not fixed. The amount of variation of the intervals is different from one interval to another.  If we look at the standard deviation of the intervals performed in various gushes in Table \ref{SD}, we notice two separate groups of intervals. The first group includes the intervals with a relatively lower standard deviation (5 < sd <8). This group includes the intervals of first tetrachord: [C-D], [D-Ek], and [Ek-F]. It also includes the interval [Bb-C] and the interval [C-D] from the higher octave. The second group of intervals have relatively higher standard deviation (11< sd <14): [F-G], [G-Ak], and [Ak-Bb], which means that these intervals are less fixed than the first group. There had been no discussion of this sort in the literature about the intervals in Iranian classical music. This study suggests that although we are generally dealing with intervals which are not fixed, we can classify the intervals based on the amount of their variations in vocal performance.

\begin{center}
\begin{table}
\centering
\footnotesize{
\begin{tabular}{ | c | c | c | c | c |}
\hline
Note Names &	Our Method & Farhat & Talāi & Vaziri \\
\hline
C &	0&	0 &	0 & 0 \\
\hline
D	&210	&205& 200 & 200\\
\hline
Ek &347	&340	&350 & 350	\\
\hline
F& 498 & 500 &500 &500	\\
\hline
G &696	&700 &700 & 700\\
\hline
Ak &836	&835	&840. & 850\\
\hline
Bb&985&995&980 & 1000\\
\hline
C &1190 &1200 &1200 & 1200\\

\hline
\end{tabular}
}
\caption{\label{shur}Scale of Shur}

\end{table}
\end{center}

\begin{center}
\begin{table}
\centering
\footnotesize{
\begin{tabular}{ | c | c | }
\hline
Interval &	Standard Deviation (cents)\\
\hline
C-D &	7.65 \\
\hline
D-Ek &	7.79\\
\hline
Ek-F & 7.07	\\
\hline
F-G& 11.47	\\
\hline
G-Ak &13.6\\
\hline
Ak-Bb &11.4\\
\hline
Bb-C&7.33\\
\hline
C-D &5.5\\

\hline
\end{tabular}
}
\caption{\label{SD}Standard Deviation of the Observed Intervals in Shur}

\end{table}
\end{center}


\section{Acknowledgements}\label{ack}

This paper is based on the author's dissertation under the supervision of Prof. Blum at the Graduate Center, City University of New York [17].

\bibliography{ISMIRtemplate}


\end{document}